\documentclass[twocolumn,amsmath,amssymb,showpacs,pre,nofootinbib,superscriptaddress]{revtex4-1} 

\bibstyle{apsrev}
\usepackage{bm}%
\usepackage{graphicx}
\usepackage{tikz}
\usepackage{ulem}
\usepackage{color}
\usepackage{xcolor}
\usepackage{physics}
\usepackage[colorlinks=true,linkcolor=blue]{hyperref}%
\expandafter\ifx\csname package@font\endcsname\relax\else
 \expandafter\expandafter
 \expandafter\usepackage
 \expandafter\expandafter
 \expandafter{\csname package@font\endcsname}%
\fi
\begin{document}

\definecolor{pyblue}{HTML}{1F77B4}
\definecolor{pyorange}{HTML}{FF7F0C}
\definecolor{pygreen}{HTML}{2CA02C}
\definecolor{pyred}{HTML}{D62728}

\title{Lattice Boltzmann simulations of stochastic thin film dewetting}

\author{S. Zitz}
\email{s.zitz@fz-juelich.de}
 \affiliation{Helmholtz Institute Erlangen-N\"urnberg for Renewable Energy,\\
  Forschungszentrum J\"ulich,\\
  F\"urther Strasse 248, 90429 N\"urnberg, Germany}%
  \affiliation{Department of Chemical and Biological Engineering, Friedrich-Alexander-Universit\"at Erlangen-N\"urnberg, F\"{u}rther Stra{\ss}e 248, 90429 N\"{u}rnberg, Germany}
\author{A. Scagliarini}%
\email{andrea.scagliarini@cnr.it}
 \affiliation{Institute for Applied Mathematics "M. Picone" (IAC), \\
Consiglio Nazionale delle Ricerche (CNR),\\
Via dei Taurini 19, 00185 Rome, Italy}%
\affiliation{INFN, sezione Roma ``Tor Vergata'', via della Ricerca Scientifica 1, 00133 Rome, Italy}
\author{J. Harting}
\email{j.harting@fz-juelich.de}
 \affiliation{Helmholtz Institute Erlangen-N\"urnberg for Renewable Energy,\\
  Forschungszentrum J\"ulich,\\
  F\"urther Strasse 248, 90429 N\"urnberg, Germany}%
 \affiliation{Department of Chemical and Biological Engineering and Department of Physics, Friedrich-Alexander-Universit\"at Erlangen-N\"urnberg, F\"{u}rther Stra{\ss}e 248, 90429 N\"{u}rnberg, Germany}
\date{\today}

\begin{abstract}
We study numerically the effect of thermal fluctuations and of variable 
fluid-substrate interactions on the spontaneous dewetting of thin liquid films.
To this aim, we use a recently developed lattice Boltzmann method for 
thin liquid film flows, equipped with a properly devised stochastic term.
While it is known that thermal fluctuations yield shorter rupture times,
we show that this is a general feature of hydrophilic substrates, irrespective of the contact angle.
The ratio between deterministic and stochastic rupture times, though, decreases with $\theta$. 
Finally, we discuss the case of fluctuating thin film dewetting on chemically patterned substrates
and its dependence on the form of the wettability gradients.

\end{abstract}

\maketitle

\newcommand{\ts}{\textsuperscript}

\section{Introduction}\label{sec:intro}
A liquid wetting a solid surface is a process that appears in many everyday life situations, like, e.g., when a rain drop hits the leaf of a plant or when coffee is spilt on a table.
Wetting phenomena have an impact on areas as diverse as surface coatings, printing, plant treatment with pesticides, or even pandemics~\cite{RevModPhys.69.931, doi:10.1146/annurev-fluid-011212-140734, cassie1944wettability, Lenormand_1990,DERYCK1998278, doi:10.1146/annurev.fluid.31.1.347,BergeronNature,BhardwajPoF2020}.
Coating processes heavily rely on the affinity between a fluid and a substrate~\cite{RevModPhys.81.739}.
If this affinity is not sufficiently strong or the substrate is rather heterogeneous the liquid layer may eventually become unstable,
leading to film rupture~\cite{RevModPhys.69.931, RevModPhys.81.1131} or wrinkling~\cite{DASILVASOBRINHO19991204}. 
These instabilities are, of course, harmful to a uniform coating.\\ 
A successful mathematical modelling of the dynamics of thin liquid films is based on the thin-film equation~\cite{Reynolds, RevModPhys.69.931}
\begin{equation}\label{eq:thin_film}
    \partial_t h(\mathbf{x},t) = \nabla\cdot\left(M(h)\nabla p(\mathbf{x},t)\right),
\end{equation}
where $h(\mathbf{x},t)$ is the height of the free surface at position $\mathbf{x} = (x,y)$ and time $t$, while $p(\mathbf{x},t)$ is the pressure at the free surface.
The function $M(h)$ denotes the mobility, whose
functional form is determined by the fluid velocity boundary condition at 
the solid surface: for a no-slip condition~\cite{RevModPhys.69.931}, in particular, $M(h)=\frac{h^3}{3\mu}$, where $\mu$ is the fluid dynamic viscosity.
The film pressure $p$ consists of the capillary pressure and the disjoining pressure, i.e.,
\begin{equation}\label{eq:pressure}
    p = - \gamma\nabla^2 h - \Pi(h),
\end{equation}
where $\gamma$ is the surface tension. The disjoining pressure
$\Pi(h)$ is the derivative, with respect to the height $h$, of the interfacial potential,
which incorporates the interactions between liquid and substrate (i.e. wetting properties).
The advantage of such approach in comparison with molecule-resolved methods such as Molecular Dynamics~\cite{haile1992molecular, PhysRevE.100.023108, doi:10.1063/1.1290698, grabow1988thin}
and Density Functional Theory~\cite{PhysRevA.40.2567, tarazona1984simple, meister1985density, hughes2014introduction} is that these 
may quickly become computationally prohibitive, as the size of the film is increased above few nanometres in thickness and to the micrometric scale in the horizontal extension.
However, when dealing with films of nanometric thickness, a hydrodynamic description may fall short due to thermal fluctuations. 
For instance, in the context of dewetting, rupture times measured in experiments are shorter than predicted by simulations of the thin-film equation~\cite{herminghaus1996dewetting,herminghaus1998dewetting,becker2003complex}.
More in general, thermal fluctuations can accelerate the appearance of film instabilities~\cite{Rauscher2008,tsekov1993effect, PhysRevLett.99.114503, PhysRevE.100.023108}.
The first theoretical and experimental studies, by means of light scattering techniques, on film rupture influenced by thermal fluctuations date back to the 60s~\cite{doi:10.1021/ja01014a015}.
Almost half a century later, a stochastic version of Eq.~(\ref{eq:thin_film}) was derived~\cite{Grun2006, Mecke_2005, PhysRevLett.95.244505}, through a lubrication approximation of the Navier-Stokes-Landau-Lifshitz 
equations of fluctuating hydrodynamics~\cite{Landau1987Fluid}.
A strong agreement between experiments and the stochastic thin-film equation was found in showing that the variance of the interfacial roughness cannot be explained without adding a fluctuating term~\cite{PhysRevLett.99.114503}.
The influence of thermal fluctuations on the short and long time morphology of a film has been studied with simulations of the one and two dimensional stochastic thin-film equations~\cite{PhysRevE.92.061002, alizadeh_pahlavan_cueto-felgueroso_hosoi_mckinley_juanes_2018}. 
Recently, new theoretical insights have been gained explaining film rupture as a consequence of the combined action of thermal fluctuations and drainage, both numerically~\cite{shah_van_steijn_kleijn_kreutzer_2019} and experimentally~\cite{PhysRevLett.125.158001}.

\textcolor{black}{To the best of our knowledge, the impact of thermal fluctuations on thin film dewetting
of chemically patterned substrates has so far been overlooked.
The purpose of this paper is to fill this gap, studying the role of heterogeneous substrate properties in the fluctuating dewetting dynamics.
To this aim we perform numerical simulations of the deterministic and stochastic thin-film equations with space-varying contact angle 
(parametrizing the chemical pattering of the substrate). In particular, since we aim to simulate large systems in order to
achieve reliable statistics in terms of the number of droplets, we restrict ourselves in this study to the one-dimensional geometry and keep the simulations computationally feasible.}

This paper is organized as follows. In Sec.~\ref{sec:stoch_thin_film} we shortly discuss the stochastic thin-film equation and its linear stability analysis. Then, we introduce the key features of our lattice Boltzmann model for 
the thin-film equation and its extension to include thermal fluctuations in Sec.~\ref{sec:num_method}. 
The results are presented and discussed in Sec.~\ref{sec:results}. 
Finally, conclusions and summary are provided in Sec.~\ref{sec:sum_conclu}. 

\section{Linear stability analysis of the stochastic thin-film equation}\label{sec:stoch_thin_film}
The one-dimensional stochastic thin-film equation reads~\cite{Grun2006, Mecke_2005}:
\begin{equation}\label{eq:stoch_thin_film_simple}
    \partial_t h = \partial_x \left[\frac{h^3}{3\mu}\partial_x p + \sqrt{\frac{2k_BTh^3}{3\mu}}\mathcal{N}\right],
\end{equation}
where $k_B$ is Boltzmann's constant, $T$ is the temperature and $\mathcal{N}$ is a Gaussian white noise with
\begin{align}\label{eq:correlation}
    \langle\mathcal{N}(x,t)\rangle =&~0,\\ 
    \quad \langle\mathcal{N}(x,t)\mathcal{N}(x',t')\rangle =&~\delta(x-x')\delta(t-t').
\end{align}
In order to understand the influence of thermal fluctuations on the temporal evolution of the film height, we cursory recall the linear stability analysis of Eq.~(\ref{eq:stoch_thin_film_simple}) 
\cite{PhysRevE.100.023108, PhysRevE.93.013120}.
Introducing the deviation $\delta h$ from the mean height $h_0$, such that $h = h_0 + \delta h$ and $\delta h \ll h_0$, 
the linearised stochastic thin-film equation is obtained
\begin{equation}\label{eq:linearstability_realspace}
    \partial_t \delta h = \frac{h_0^3}{3\mu}\Big(-\Pi^{\prime}(h_0)\partial_x^2\delta h - \gamma \partial_x^4\delta h\Big) + \sqrt{\frac{2k_BT h_0^3}{3\mu}} \partial_x \mathcal{N},
\end{equation}
where $\Pi^{\prime}(h_0) \equiv \frac{\partial \Pi(h)}{\partial h}\bigg\rvert_{h=h_0}$.
Here, it is assumed that also the noise is small, in the sense that $k_B T \ll \gamma h_0^2$.
To derive a dispersion relation for this system, first the Fourier transforms of the perturbation $\delta h$  
\begin{equation}\label{eq:fourier_delta_h}
    \tilde{\delta h}(q,t) = \int_{-\infty}^{\infty} \delta h(x,t) e^{-ixq} dx,
\end{equation}
and the noise term $\mathcal{N}$
\begin{equation}\label{eq:fourier_noise}
    \tilde{\mathcal{N}}(q,t) = \int_{-\infty}^{\infty} \mathcal{N}(x,t) e^{-ixq} dx.
\end{equation}
must be performed.

Inserting $\tilde{\delta h}$ and $\tilde{\mathcal{N}}$ in~(\ref{eq:linearstability_realspace}) yields
\begin{equation}\label{eq:stability_fourier}
    \partial_t \tilde{\delta h} = \omega(q)\tilde{\delta h} +i\sqrt{\frac{2k_BTh_0^3}{3\mu}}q\tilde{\mathcal{N}},
\end{equation}
with dispersion relation $\omega(q)$ given by
\begin{equation}\label{eq:dispersion}
    \omega(q) = \frac{1}{t_0}\left[2\left(\frac{q}{q_0}\right)^2 - \left(\frac{q}{q_0}\right)^4\right].
\end{equation}
The maximum, $q_0$, of $\omega(q)$ and the characteristic time, $t_0$, are specific to the substrate the film is deposited on and their expressions are~\cite{PhysRevLett.99.114503}
\begin{equation}\label{eq:q0}
    q_0^2 = \frac{1}{2\gamma}\frac{\partial \Pi(h)}{\partial h}\bigg\rvert_{h=h_0},
\end{equation}
and
\begin{equation}\label{eq:t0}
    t_0 = \frac{3\mu}{\gamma h_0^3 q_0^4}.
\end{equation}
Solving Eq.~(\ref{eq:stability_fourier}) leads to the following expression for the spectrum~\cite{PhysRevE.100.023108,Mecke_2005},
$S(q,t) = \langle |\tilde{\delta h}(q,t)|^2\rangle$ (where the brackets stand for the average over the noise):
\begin{equation}\label{eq:structure_factor}
    S(q,t) = S_0(q)e^{2\omega(q)t} + \frac{k_BTh_0^3}{3\mu}\frac{q^2}{\omega(q)}(e^{2\omega(q)t} - 1).
\end{equation}
For $q>\sqrt{2}q_0$, $S(q,t)$ tends to the capillary wave spectrum, 
$S \sim S_{\text{cw}} \propto \frac{k_BT}{\gamma q^2}$, as $t \rightarrow \infty$ (whereas it would decay exponentially to zero in the deterministic case) ~\cite{PhysRevLett.99.114503,Mecke_2005}.
Moreover, while the maximum $q_0$ is independent of time in the deterministic case, the maximum $q_m$ of 
(\ref{eq:structure_factor}) approaches $q_0$ from the right as the time increases.
It follows that the most unstable wavelength, $\lambda_{\text{max}} = 2\pi/q_m$, grows with time.
We will come back to this fact in Sec.~\ref{sec:results} and show that indeed this behavior is reproduced and observed in our simulations.


\section{Thin Film Lattice Boltzmann Model}\label{sec:num_method}
Performing numerical simulations of the thin-film equation is challenging, even in the deterministic case;
including stochastic terms adds a further level of complexity.
To this aim, several sophisticated numerical methods have been developed, based on, e.g.,   
finite differences~\cite{PhysRevE.63.011208}, finite elements~\cite{Grun2006} and spectral schemes~\cite{Duran_Olivencia2019} . 

\subsection{Lattice Boltzmann Method}\label{subsec:LBM}
To simulate the dynamics of thin liquid films described by Eq.~(\ref{eq:thin_film}), we employ a recently developed~\cite{PhysRevE.100.033313} lattice Boltzmann method (LBM), built on 
a class of models originally devised for the shallow water equations~\cite{Salmon:1999:0022-2402:503,PhysRevE.65.036309,zhou2004lattice,van2010study}.
The one-dimensional lattice Boltzmann equation for the discrete particle distribution functions $f_i$ reads
\begin{equation}\label{eq:LBE}
    \begin{split}
        &f_i(x+c^{(i)}\Delta t,t+\Delta t) = \\
        &\left(1 - \frac{\Delta t}{\tau}\right) f_i(x,t) + \frac{\Delta t}{\tau} f_i^{(eq)}(x,t) + w_i \frac{\Delta t}{c_s^2} c^{(i)} F,
    \end{split}
\end{equation}
where $i$ labels the lattice velocities $c^{(i)}$ and runs from $0$ to $Q-1$, with $Q$ being the number of velocities characterizing the scheme, and $F$ is a (generalised) force\footnote{Strictly speaking its dimensions are 
$[\text{length}]^2[\text{time}]^{-2}$}.
Algorithmically, this equation is composed of two steps.
In the local collision step, the distribution functions $f_i$ relax towards their local equilibria $f^{(eq)}_i$ with a relaxation time $\tau$.
In the streaming step, the $Q$ distribution functions $f_i$ are moved over the lattice according to their lattice velocities $c^{(i)}$.
We adopt here the standard one-dimensional D1Q3 scheme with $3$ lattice velocities, given by
\begin{equation}\label{eq:speeds}
c^{(i)}  = [0, c, -c], \quad i = 0, 1, 2,
\end{equation}
where $c=\frac{\Delta x}{\Delta t}$, with weights
\begin{equation}
w_0 = \frac{2}{3},\quad w_{1,2} = \frac{1}{6},
\end{equation}
and speed of sound $c_s^2=\frac{c^2}{3}$.
The equilibrium distribution functions $f_i^{(eq)}$ read~\cite{van2010study}:
\begin{gather}
    f_{0}^{eq} = h\left(1-\frac{1}{2c^2}gh - \frac{1}{c^2}u^2\right),\nonumber\\
    f_{1}^{eq} = h\left(\frac{1}{4c^2}gh + \frac{1}{2c}u + \frac{1}{2c^2}u^2\right)\label{eq:equilibria},\\
    f_{2}^{eq} = h\left(\frac{1}{4c^2}gh - \frac{1}{2c}u + \frac{1}{2c^2}u^2\right),\nonumber
\end{gather}
with $g$ being the gravitational acceleration (which will be 
set to zero, hereafter, as it can be obviously neglected for nanoscale thin films).
The hydrodynamic fields (film height $h$ and velocity $u$) are expressed in terms of the 
distribution functions $f_i$ as:
\begin{equation}\label{eq:hydrofields}
h= \sum_{i=0}^2 f_i \qquad hu = \sum_{i=0}^2 c^{(i)} f_i.
\end{equation}
The force $F$ in (\ref{eq:LBE}) consists of two terms,
\begin{equation}\label{eq:force}
F = F_{\text{cap}} + F_{\text{fric}}.  
\end{equation}
In the first term the effect of the film pressure $p$, Eq.~(\ref{eq:pressure}), is 
embedded, i.e. 
\begin{equation}\label{eq:capillary_force}
    F_{\text{cap}} = -\frac{1}{\rho_0} h \frac{\partial p}{\partial x},
\end{equation}
($\rho_0$ being the fluid density) whereas the second one represents a viscous friction with the substrate:
\begin{equation}\label{eq:fric_force}
    F_{\text{fric}} = -\nu \alpha_{\delta}(h) u,
\end{equation}
where $\nu=\mu/\rho_0$ is the fluid kinematic viscosity (related to the relaxation time $\tau$ by $\nu = c_s^2\left(\tau-\frac{\Delta}{2}\right)$) and the friction factor
\begin{equation}\label{eq:fric_alpha}
     \alpha_{\delta}(h) = \frac{6 h}{2h^2 + 6h\delta + 3\delta^2},
\end{equation}
depends on the film height and on the particular fluid velocity boundary 
condition at the substrate surface, parameterised by a slip length $\delta$.
It can be shown that in the limit of vanishing inertia, negligible gravity and small film thickness, the model defined by Eqs.~(\ref{eq:LBE}-\ref{eq:fric_alpha})
represents a numerical solver for the system of equations
\begin{equation}\label{eq:lubr2eq}
\begin{cases}
\begin{array}{ll}
\partial_t h + \partial_x (h u)  = 0 & \\ 
\partial_t (h u) = -\frac{1}{\rho_0}h\partial_x p -\nu\alpha_{\delta}(h)u.
\end{array}
\end{cases}
\end{equation}
The left hand side of the second of Eqs.~(\ref{eq:lubr2eq}) can be neglected in the limits considered (small Reynolds number and film thickness). Thus, 
the equation reduces to $u \approx -\frac{h}{\mu\alpha_{\delta}(h)}\partial_x p$ that, plugged into the first of Eqs.~(\ref{eq:lubr2eq}), gives Eq.~(\ref{eq:thin_film}) 
(with a no-slip mobility $M(h)=\frac{h^3}{3\mu}$ if the further limit $\delta \rightarrow 0$ is taken).

\subsection{Modelling Thermal Fluctuations}\label{subsec:thermal_fluc_model}
In order to enable our method to simulate the stochastic thin-film equation, we proceed in a constructive way,
introducing a fluctuating force, $F_{\text{fluc}}$, in the velocity equation in (\ref{eq:lubr2eq}) such that, in the limits previously discussed, Eq.~(\ref{eq:stoch_thin_film_simple}) is recovered. The velocity entering the height equation in (\ref{eq:lubr2eq}) 
then reads
\begin{equation}
u \approx -\frac{h}{\mu \alpha_{\delta}(h)}\partial_x p + \frac{\rho_0}{\mu \alpha_{\delta}(h)}F_{\text{fluc}}.
\end{equation}
A term-by-term matching with Eq.~(\ref{eq:stoch_thin_film_simple}) straightaway tells that the sought form for the fluctuating force is
\begin{equation}
    F_{\text{fluc}} = \frac{1}{\rho_0} \sqrt{2 k_B T \mu \alpha_{\delta}(h)}\mathcal{N}
\end{equation}
(where we have omitted the sign since $\mathcal{N}$ is a zero-mean random number), or also, in the no-slip 
limit $\delta \rightarrow 0$, 
\begin{equation}\label{eq:thermal_force}
    F_{\text{fluc}} = \frac{1}{\rho_0}\sqrt{\frac{6k_BT\mu}{h}}\mathcal{N}.
\end{equation}
The fluctuating force $F_{\text{fluc}}$ enters in the scheme by being added to the total force, Eq.~(\ref{eq:force}), that becomes
\begin{equation}\label{eq:tot_force}
F = F_{\text{cap}} + F_{\text{fric}} + F_{\text{fluc}}
\end{equation}
and appears in the lattice Boltzmann equation (\ref{eq:LBE}).

\subsection{Fluid-substrate interactions: the contact angle}\label{subsec:fluid_substrate}
For the disjoining pressure $\Pi(h)$ we use the following 
expression\textcolor{black}{\cite{RevModPhys.81.1131, Peschka9275, PhysRevE.92.061002, RevModPhys.69.931}}:
\begin{equation}\label{eq:disjoin_p}
    \Pi(h) = \kappa(\theta) \left[\left(\frac{h_{\ast}}{h}\right)^n - \left(\frac{h_{\ast}}{h}\right)^m\right],
\end{equation}
where the prefactor $\kappa(\theta) = \gamma(1-\cos(\theta))\frac{(n-1)(m-1)}{(n-m)h_{\ast}}$ is related to the Hamaker constant by
$\mathcal{A} = 6 \pi h_{\ast}^3 \kappa(\theta)$.
$\theta$ is the contact angle, which effectively encodes the interfacial energies of the three-phase system and represents, therefore, a measure of the hydrophilicity/hydrophobicity of the substrate.
\textcolor{black}{Throughout the paper we restrict ourselves to low contact angles, $\theta < 1$,
so to comply with the lubrication approximation, which requires the derivative of the height field to be small, $\partial_x h(x) \ll 1$, and hence limits
the values of $\theta$ (since, at the contact point, $\partial_x h \sim \tan(\theta) \sim \theta$).}
The value $h_{\ast} > 0$, 
at which the disjoining pressure vanishes, sets the thickness of the precursor film covering the substrate.
\section{Results}\label{sec:results}
\begin{center}
\begin{table}
    \begin{tabular}{||l | c | c | p{1cm} | c | c | p{1cm} ||}
        \hline
        & \multicolumn{3}{ | c | }{$n=3, m=2$} & \multicolumn{3}{ | c | }{$n=9, m=3$} \\
        \hline
         $\theta$ & $t_0$ & $q_0$ & $\lambda_{\text{max}}$ & $t_0$ & $q_0$ & $\lambda_{\text{max}}$ \\ [0.5ex] 
         \hline\hline
         $\pi/9$ & $5 \cdot 10^5$ & $0.101$ & $62$ & $8.6 \cdot 10^6$ & $0.049$ & $128$ \\ 
         \hline
         $5\pi/36$ & $2 \cdot 10^5$ & $0.126$ & $50$ & $3.5 \cdot 10^6$ & $0.061$ & $103$  \\
         \hline
         $\pi/6$ & $10^5$ & $0.151$ & $42$ & $1.7 \cdot 10^6$ & $0.073$ & $86$ \\
         \hline
         $7\pi/36$ & $5 \cdot 10^4$ & $0.175$ & $36$ & $10^6$ & $0.085$ & $74$ \\
         \hline
         $2\pi/9$ & $3 \cdot 10^4$ & $0.2$ & $32$ & $6 \cdot 10^5$ & $0.097$ & $65$ \\ [1ex] 
         \hline
        \end{tabular}
        \caption{Numerical values (in lbu) of characteristic time, $t_0$, most unstable wavenumber, $q_0$, and corresponding wavelength, $\lambda_{\text{max}}$, 
        for the various contact angles $\theta$ and pairs $(n,m)$ of the disjoining pressure exponents used (see Eq.~(\ref{eq:disjoin_p}).}
    \label{tab:t_0s}
\end{table}
\end{center}
\textcolor{black}{We numerically solve the model 
(\ref{eq:LBE}-\ref{eq:hydrofields}), with forcing term (\ref{eq:capillary_force}-\ref{eq:fric_force},\ref{eq:thermal_force}-\ref{eq:tot_force}) (and $\Delta x = \Delta t = 1$), on a domain 
of length $L=4096$,
with a relaxation time $\tau = 1$, corresponding to a kinematic viscosity $\nu = 1/6$.
The surface tension $\gamma$ is set to $0.01$ and we use, as exponents appearing in the disjoining pressure (Eq. (\ref{eq:disjoin_p})),
the pairs $(n,m)=(3,2)$ (in section \ref{subsec:validation}, \ref{subsec:morphandrup} and \ref{subsec:results_contact_angle}) 
and $(n,m)=(9,3)$ (in section \ref{subsec:patterned_substrates}).
The thickness of the precursor film is $h_{\ast}= 0.05$ and the numerical slip length $\delta$ (unless otherwise stated)
is chosen to be in the weak slip regime, $\delta/h_0 = 1.0$. 
The system is initialised with zero velocity everywhere ($u(x,0)=0 \; \forall x \in [0, L]$)
and with random perturbations of amplitude $10^{-3}$ around the value $h_0 = 1$ for the film height, i.e. 
$h(x,0) = h_0 + 10^{-3} \varepsilon$ ($\varepsilon$ being a random variable uniformly distributed 
in $[-1, +1]$).
In the remainder of the manuscript all lengths will be given adimensionalised by $h_0$.
We vary the contact angle in the interval $[\pi/9, 2\pi/9 ]$;
the numerical values (in lattice Boltzmann units, lbu) of the corresponding wavelength, $\lambda_{\text{max}}$ and wavenumber, $q_0$, of the most unstable mode as well as the characteristic time, $t_0$, can be found in Tab.~\ref{tab:t_0s}.
For each contact angle configuration we perform one simulation for the athermal system and a set 
of $O(30)$ runs for the fluctuating case, at changing the random seed (the data shown represent, then, an average over the noise realisations). We would like to stress, incidentally, that the inclusion of 
the stochastic term entails a computational overhead of, roughly, $20\%$ of a typical deterministic run.\\ 
The numerical data will be indicated, hereafter, in terms of the dimensionless temperature $\sigma = \sqrt{\frac{k_B T}{\gamma h_0^2}}$, which is set either to $\sigma=0$ (for the athermal dewetting) or to $\sigma = 3.16 \cdot 10^{-3} \equiv \sigma_0$, 
corresponding to a thermal energy
of $k_B T = 10^{-7}$ lbu (for the fluctuating dewetting). For the sake of comparison, we remark 
that such a value lies well within the window $\sigma \in [10^{-3}, 10^{-2}]$, measured in experiments with polymeric and metallic thin films~\cite{PhysRevLett.99.114503,doi:10.1021/la4009784}}.

\subsection{Testing the stochastic term}\label{subsec:validation}
\begin{figure}
    \centering
    \includegraphics[width=0.48\textwidth]{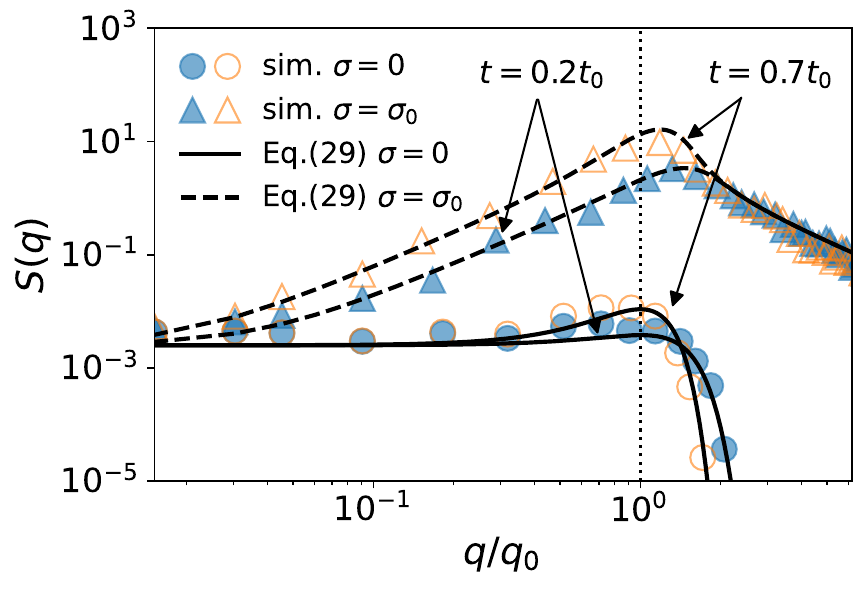}
    \caption{\textcolor{black}{(color online)} Height fluctuations spectra from deterministic (circles) and stochastic (triangles) simulations at $t=0.2 t_0$ (filled symbols) and $t=0.7 t_0$ (empty symbols), 
    on a substrate with $\theta =\pi/9$. The theoretical predictions, Eq.~(\ref{eq:structure_factor_2}), are reported with solid ($\sigma=0$) and dashed ($\sigma = \sigma_0$) lines.
    }
    \label{fig:theory_simulation_structure_factor}
\end{figure}
\textcolor{black}{In order to validate our model against the analytical results discussed
in section \ref{sec:stoch_thin_film}, we first rewrite Eq.~(\ref{eq:structure_factor_2}) in a form, involving the dimensionless temperature $\sigma$, which is more 
convenient for the forthcoming discussion, namely:
\begin{equation}\label{eq:structure_factor_2}
    S(q,t) = S_0 e^{2\omega(q)t} + \frac{\sigma^2 h_0^2 L}{q_0^2}g\left(\frac{q}{q_0}\right)(e^{2\omega(q)t} - 1),
\end{equation}
where $g(\xi) = \frac{\xi^2}{2\xi^2 - \xi^4}$ and the appearance of the system size $L$ stems from the noise amplitude, $\langle |\tilde{\mathcal{N}}(q,t)|^2 \rangle = L$, 
resulting from discrete Fourier-transforming 
over a finite length~\cite{PhysRevE.100.023108,PhysRevE.102.053105}. To arrive at (\ref{eq:structure_factor_2}), use was made of the expression (\ref{eq:t0}) for the characteristic time $t_0$
and we omitted the dependence on $q$ of $S_0(q)$ because, with our initialization, the latter is just a constant, $S_0(q)\equiv S_0$.}
We measured, then, the spectra $S(q,t)$, at two instants of time, 
in the early stages of the growth of the instability, for both athermal and fluctuating dewetting on a susbstrate with $\theta=\pi/9$ (and strict no-slip, $\delta=0$).
The data, reported in Fig.~\ref{fig:theory_simulation_structure_factor}, from both deterministic (circles)
and fluctuating (triangles) simulations show good agreement with the theoretical curves, Eq.~(\ref{eq:structure_factor_2}), depicted with solid ($\sigma=0$) and dashed ($\sigma=\sigma_0$) lines.
In particular, notice that the maximum  of $S(q,t)$ is attained for $q=q_0$ independently of time, in the athermal
case, and at $q = q_m > q_0$, with $q_m$ tending to $q_0$ as the time goes by, when 
fluctuations are present; in the latter case, in addition, the capillary wave spectrum $S_{\text{cw}}(q) \propto q^{-2}$
is recovered for large $q$.
\subsection{Droplet size distributions}\label{subsec:morphandrup}
\begin{figure}
    \centering
    \includegraphics[width=0.48\textwidth]{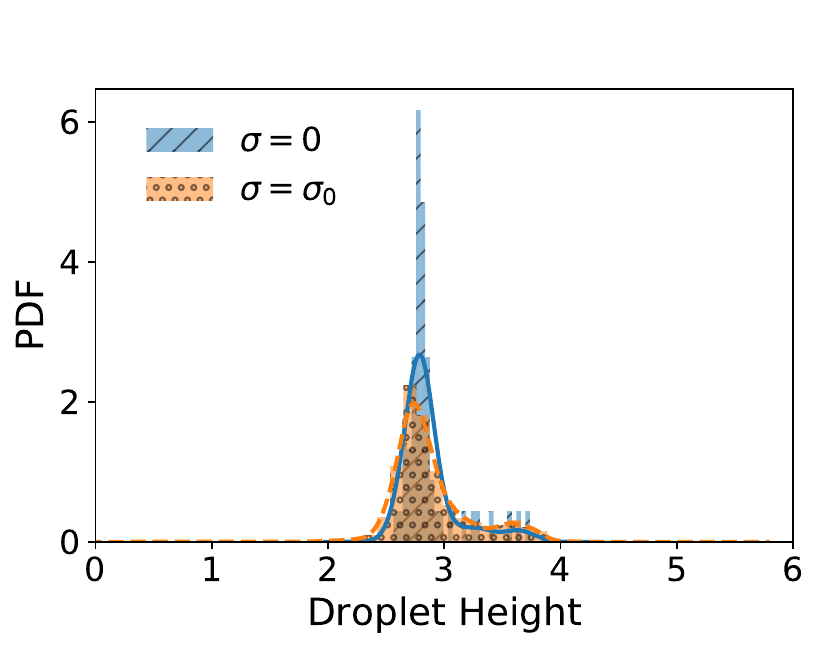}
    \caption{\textcolor{black}{(color online)} Histograms of the droplet height distributions from the deterministic (blue, line-patterned) and stochastic (orange, dot-patterned) simulations with 
    $\theta = \pi/9$, at $t \approx 20 t_0$; the corresponding Gaussian kernel density estimations are also plotted, as a guide to the eye, with solid blue (dashed orange) curve for the deterministic (stochastic) data.
    }
    \label{fig:droplet_distribution}
\end{figure}
In the early stage of the deterministic dewetting, we expect that 
the droplet size distribution, quantified in 
terms of the droplet height, will be strongly correlated to the maximum unstable wavelength 
$\lambda_{\text{max}} =2\pi/q_0$:
\begin{equation}\label{eq:lambda_max}
    \lambda_{\text{max}} = \frac{2\pi}{q_0} = \sqrt{\frac{8\pi^2\gamma}{\Pi'(h)|_{h_0}}} 
\end{equation}
More quantitatively, in our one-dimensional case the equilibrium shape of a droplet is a circular arc, 
whose chord is the portion of droplet in contact with the substrate and equals $\lambda_{\text{max}}/2$. Therefore, it can be 
shown, by means of simple geometrical arguments, that the droplet height is
\begin{equation}\label{eq:cap_height}
   h_{\text{drop}} = \frac{\lambda_{\text{max}}}{4}\tan\left(\frac{\theta}{2}\right), 
\end{equation}
where $\theta$ is the contact angle. For $\theta = \pi/9$, we get 
$\lambda_{\text{max}} \approx 62$ and $h_{\text{drop}} \approx 2.7$, which is the maximum around which the height values should be distributed.
This is, indeed, what we 
observe in Fig.~\ref{fig:droplet_distribution}, where we show measurements  
of the droplet height distribution obtained from simulations with and without thermal fluctuations.
In the deterministic case ($\sigma=0$), the distribution (blue, line-patterned, histogram) sharply peaks 
around the theoretically predicted value $h_{\text{drop}} \approx 2.7$. 
As a guide to the eye we add a Gaussian kernel density estimation (solid blue curve) that shows the best fitting continuous distribution to 
the histogram. 
The data from the stochastic simulations ($\sigma = \sigma_0$), instead, show a broadening of 
the distribution (orange, dot-patterned, histogram and dashed orange curve).
Adding thermal energy, then, on the one hand, facilitates the system to explore higher energetic states, whereas
on the other hand it reduces the coarsening time scales~\cite{Grun2006}: this is reflected in the tails of the distribution for 
both small and large values of $h$, respectively.
These findings are in agreement with what was reported by Nesic et al.~\cite{PhysRevE.92.061002}.
\begin{figure}
    \centering
    \includegraphics[width = 0.48\textwidth]{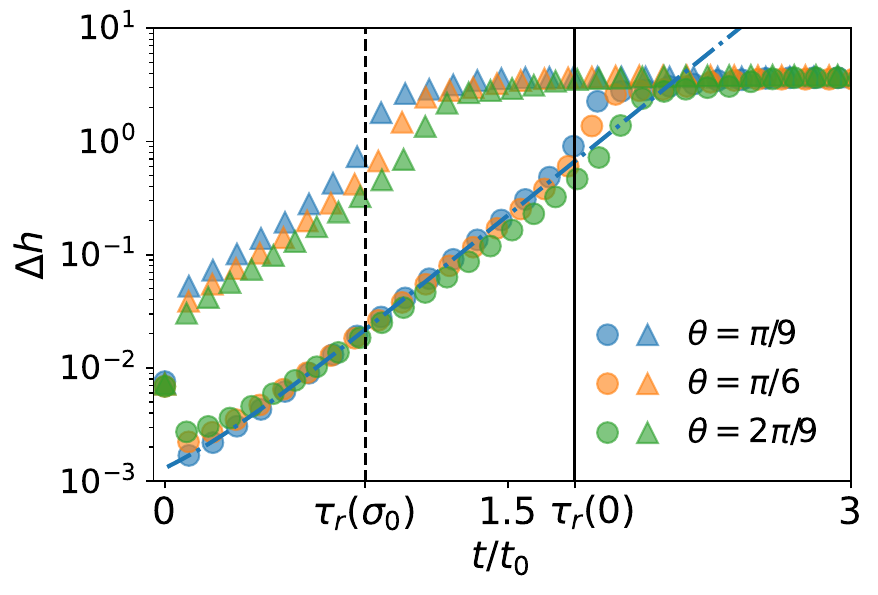}
    \caption{\textcolor{black}{(color online)} Time evolution of $\Delta h(t) = \max_x\{h(x,t)\} - \min_x\{h(x,t)\}$ for the athermal (bullets) and fluctuating (triangles) systems with $\theta=\pi/9$.
             The blue dashed-dotted line depicts an exponential fit of the data for the athermal case ($\sigma=0$).
             The vertical lines mark the rupture times for the deteministic (solid) and stochastic (dotted) simulations.}
    \label{fig:delta_heigth_evo}
\end{figure}

\subsection{Rupture times and role of contact angle}\label{subsec:results_contact_angle}
\begin{figure}
    \centering
    \includegraphics[width=0.48\textwidth]{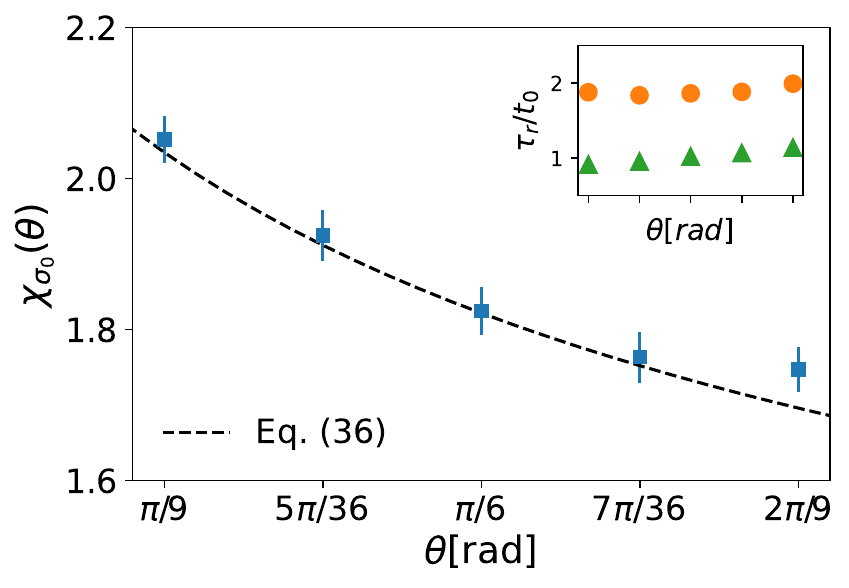}
    \caption{\textcolor{black}{(color online) Ratio of rupture times from athermal and fluctuating 
    dewetting, $\chi_{\sigma_0}(\theta)$ (Eq.~(\ref{eq:defchi}), as a function of the contact angle 
    $\theta$. The dashed line is Eq.~(\ref{eq:chi}) for $\sigma=\sigma_0$.
    Inset: Rupture times $\tau_r$ normalized by $t_0$ (see Tab.~\ref{tab:t_0s}) vs $\theta$, 
    for the deterministic (orange bullets \textcolor{pyorange}{$\bullet$}) and stochastic 
    (green triangles  \textcolor{pygreen}{$\blacktriangle$}) simulations.}
    }
    \label{fig:rupture_times_semilogy_more_theta}
\end{figure}

\begin{figure}
    \centering
    \includegraphics[width=0.48\textwidth]{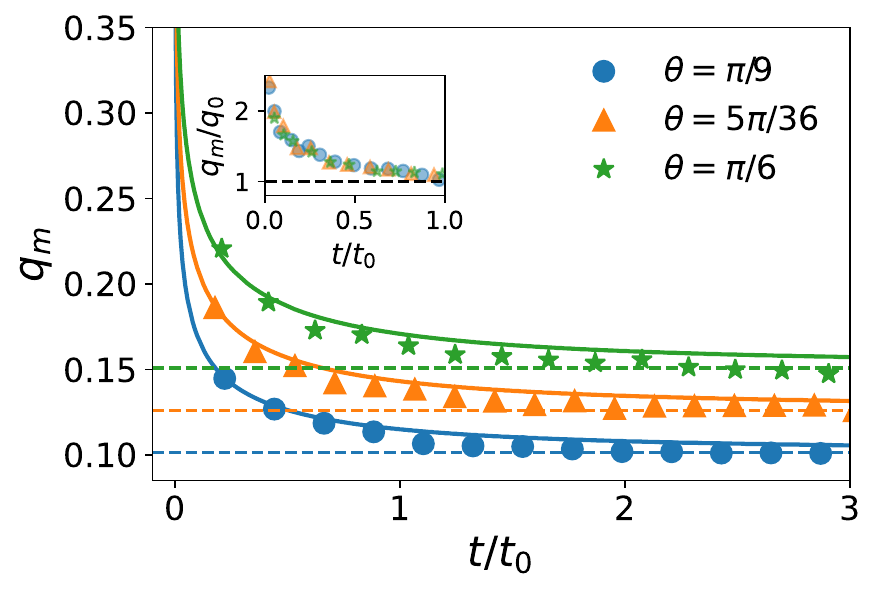}
    \caption{\textcolor{black}{(color online)} Time dependence of the maximum wavenumber $q_m$ for various contact angles $\theta$ and $\sigma = \sigma_0$, using $\delta=0$.
    The full blue, orange and green lines correspond to theoretical curve $q_m(t)$ derived from Eq.~(\ref{eq:structure_factor_2}) for $\theta = \pi/9, 5\pi/36, \pi/6$.
    The dashed lines show the corresponding $q_0(\theta)$. 
    In the inset we show that the data collapse onto a single curve upon rescaling each $q_m$ by its respective $q_0$.}
    \label{fig:evolution_spectrum_peak_with_t}
\end{figure}

Let us focus, now, on the dependence of the rupture times of the spinodally dewetting film on temperature and contact angle. 
To this aim, we compare, in Fig.~\ref{fig:delta_heigth_evo}, the evolution in time of height perturbations, $\Delta h(t) = \max_x\{h(x,t)\} - \min_x\{h(x,t)\}$, from deterministic (bullets (\textcolor{pyblue}{$\bullet$}, \textcolor{pyorange}{$\bullet$}, \textcolor{pygreen}{$\bullet$})) and stochastic (triangles (\textcolor{pyblue}{$\blacktriangle$}, \textcolor{pyorange}{$\blacktriangle$}, \textcolor{pygreen}{$\blacktriangle$})) simulations, for three different contact angles, $\theta = \{ \pi/9, \pi/6, 2\pi/9 \}$ (color-coded).
A similar analysis can also be found in Ref.~\cite{alizadeh_pahlavan_cueto-felgueroso_hosoi_mckinley_juanes_2018}.
As predicted by the linear stability analysis (see Eq~(\ref{eq:stability_fourier})), the perturbations initially follow an exponential growth law (highlighted by the dashed-dotted blue line, fitting the deterministic data for $\theta=\pi/9$);
at later times, the data start to deviate from the linear stability 
prediction and a kink in the curves appears, signalling the film rupture. As expected, in the fluctuating case the rupture occurs earlier. \\
\textcolor{black}{A deeper insight can be achieved looking at the dependence of the rupture events on the contact angle. Defining the rupture time $\tau_r$ as 
the earliest instant of time at which the free surface ``touches'' the substrate
$$
\tau_r = \min\{t|h(x,t) = h_{\ast}\},
$$
we find that with thermal fluctuations ($\sigma \neq 0$), $\tau_r$ is shorter than the athermal counterpart.
The latter feature holds true irrespective of the contact angle, i.e. 
$$
\tau_r^{(\sigma=\sigma_0)}(\theta) < \tau_r^{(\sigma=0)}(\theta) \quad \forall \theta,
$$
as it can be better appreciated from Fig.~\ref{fig:rupture_times_semilogy_more_theta}, where we plot 
the ratio
\begin{equation}\label{eq:defchi}
\chi_{\sigma_0}(\theta) \stackrel{\text{def}}{=} \frac{\tau_r^{(\sigma=0)}(\theta)}{\tau_r^{(\sigma=\sigma_0)}(\theta)}
\end{equation}
as a function of $\theta$ and we see that it is always larger than one. 
Remarkably, moreover, while in the athermal system the ratio $\tau_r^{(\sigma=0)}(\theta)/t_0$ is basically independent of $\theta$, in the fluctuating case
a weak growth of $\tau_r^{(\sigma=\sigma_0)}(\theta)/t_0$ can be detected (see inset of 
Fig.~\ref{fig:rupture_times_semilogy_more_theta}).
Consequently, $\chi_{\sigma_0}(\theta)$ is a (monotonically) decreasing function of  
the contact angle.\\
These findings can be understood as follows. The rupture time is such that $\Delta h(\tau_r) \sim h_0$; if we assume 
the validity of the linear regime (or, equivalently, an exponential evolution) up to rupture 
(as it looks reasonable from Fig.~\ref{fig:delta_heigth_evo}), we can estimate $\Delta h(t)$ as 
$$
\Delta h(t) \sim \left(\frac{S(q_m,t)}{L}\right)^{1/2}
$$
(where it is also assumed that the growth rate is dominated by the maximum of the spectrum, $q_m$\footnote{The factor $1/L$ ensures the 
fulfillment of the discrete Parseval-Plancherel's theorem}).
At the rupture time we have
\begin{equation}\label{eq:ruptime}
\left(\frac{S(q_m,\tau_r)}{L}\right)^{1/2} \sim h_0.
\end{equation}
In the deterministic case $q_m=q_0$ and $S(q_0,t) = S_0e^{2t/t_0}$, 
whence
\begin{equation}\label{eq:taur_det}
    \tau_r^{(\sigma=0)} \sim t_0 \log\left(\frac{h_0 L^{1/2}}{S_0^{1/2}}\right).
\end{equation}
In the stochastic case, the time dependence of $S(q_m,t)$ is slightly more complicated, but it can be simplified noticing that, 
when $t \sim t_0$ (i.e. close to rupture), $q_m \approx q_0$ (see Fig.~\ref{fig:evolution_spectrum_peak_with_t}) and, therefore, 
in Eq.~(\ref{eq:structure_factor_2}), $g(q_m/q_0) \approx g(1) = 1$ and $(e^{2\omega(q_m)t}-1) \approx (e^{2t/t_0}-1) \approx e^{2t/t_0}$.
The spectrum, then, reduces to 
$$
S(q_m,t) \approx \left(S_0 + \frac{\sigma^2 h_0^2 L}{q_0^2}\right)e^{2t/t_0},
$$
which, by virtue of (\ref{eq:ruptime}) and neglecting $S_0$ (for $S_0 \ll \frac{\sigma^2 h_0^2 L}{q_0^2}$), delivers
\begin{equation}\label{eq:taur_stoch}
    \tau_r^{(\sigma)} \sim t_0 \log\left(\frac{q_0}{\sigma}\right) \approx t_0 \log \left( \frac{a \theta}{\sigma}\right).
\end{equation}
Here, we used the expression for the deterministic maximum wavenumber $q_0$, i.e. (from Eqs.~(\ref{eq:q0}) and (\ref{eq:disjoin_p})) 
$$
\!q_0^2\!=\!\frac{1-\cos \theta}{2h_{\ast}^2(n\!-\!m)}(n\!-\!1)(m\!-\!1)\!\left[m\left(\frac{h_{\ast}}{h_0}\right)^{m+1}\! -\! n\left(\frac{h_{\ast}}{h_0}\right)^{n+1}\right]\!,
$$ 
which can, in the small angle approximation $1-\cos \theta \approx \frac{\theta^2}{2}$, 
be written as $q_0^2 \approx a^2 \theta^2$.
Taking the ratio of (\ref{eq:taur_det}) and (\ref{eq:taur_stoch}) we get
\begin{equation}\label{eq:chi}
    \chi_{\sigma}(\theta) \sim \frac{\log\left(\frac{h_0 L^{1/2}}{S_0^{1/2}}\right)}{\log\left(\frac{a \theta}{\sigma}\right)} \propto \frac{1}{\log(\theta/\sigma)}.
\end{equation}
The latter relation tells us that $\chi_{\sigma}(\theta)$, indeed, increases with the temperature 
(confirming that the rupture times are shorter for the fluctuating systems), but decreases with the contact angle.}
\textcolor{black}{Eq.~(\ref{eq:chi}) is plotted in Fig.~\ref{fig:rupture_times_semilogy_more_theta} (dashed line), showing a nice agreement with the numerical data.}\\
Fig.~\ref{fig:evolution_spectrum_peak_with_t} depicts the evolution of the maximum wavenumber $q_m$ with time for different values of the contact angle.
The solid lines show the theoretical curves $q_m(t)$ derived from Eq.~(\ref{eq:structure_factor_2}).
In compliance with the theory exposed in \ref{sec:stoch_thin_film}, we see that, for each $\theta$, $q_m$ decreases in time and tends asymptotically to the corresponding $q_0(\theta)$, 
indicated with dashed lines.

\subsection{Patterned substrates}\label{subsec:patterned_substrates}
On patterned substrates one generally observes that a fluid prefers to wet regions with lower contact angle,
thus the modulated wettability induces a net force (effectively entering in our model through a space-varying contact angle, $\theta=\theta(x)$, 
in Eq.~(\ref{eq:capillary_force})).
\textcolor{black}{We will now show how simple patterns can substantially change the droplet distribution, with and without thermal fluctuations, thus allowing, in principle, to control both the size and the number of droplets generated during the dewetting process.}
We consider two kinds of patterns: a sine wave and a square wave (the latter mimicking the effect of spatially confined defects that 
induce contact angle "jumps").
In the following we show data obtained using the disjoining pressure exponents $(9,3)$ instead of $(3,2)$.
The reason for this choice is that although the overall behavior is the same, the characteristic time scale $t_0$ is larger, as indicated in Tab.~\ref{tab:t_0s},
thus extending the duration of the height field evolution and, therefore, allowing for a much clearer presentation of the data.

\subsubsection{Sine wave pattern}\label{subsubsec:sine}
\begin{figure}
    \centering
    \includegraphics[width=0.48\textwidth]{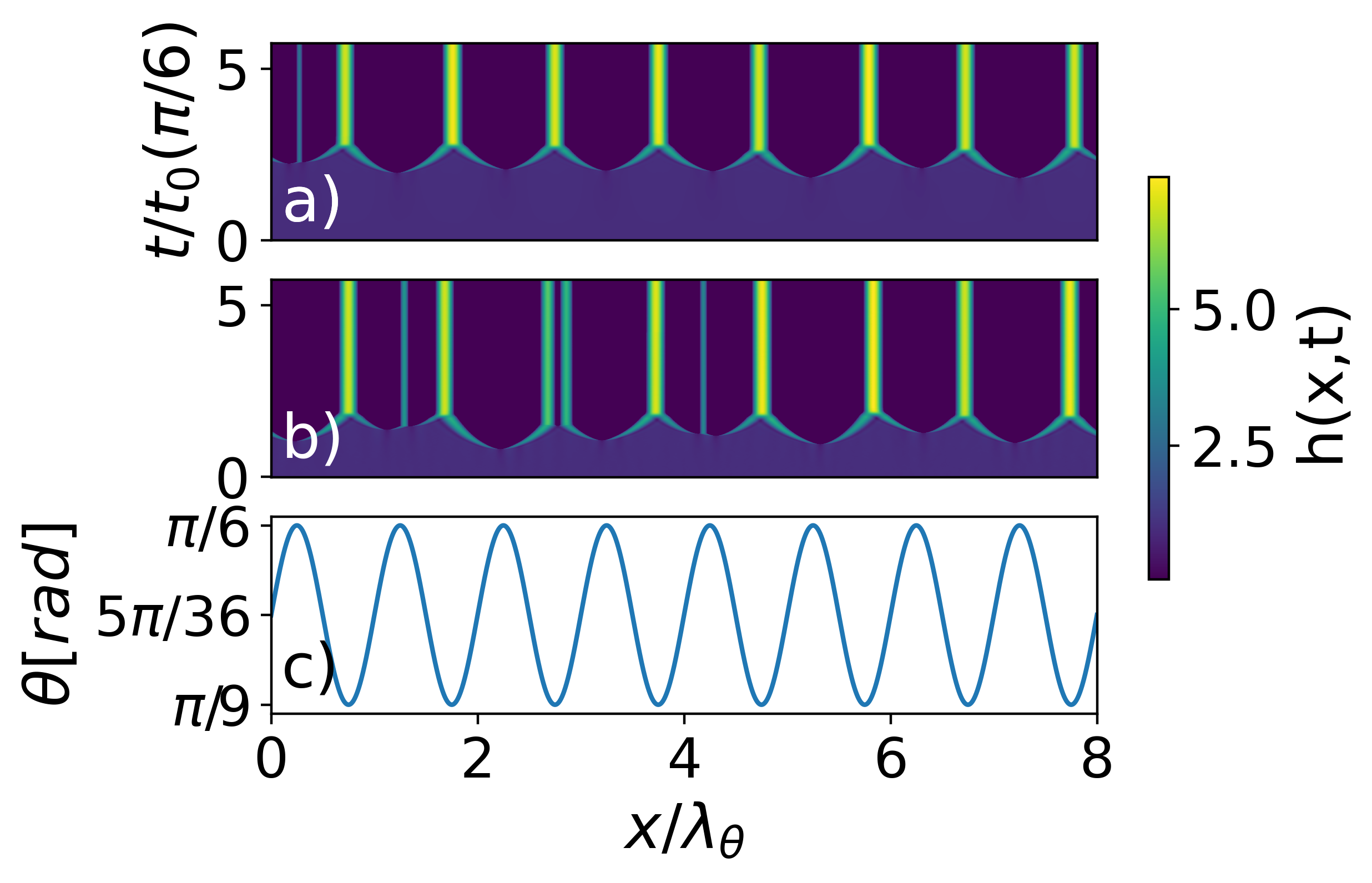}
    \caption{\textcolor{black}{(color online)} Space-time plot of the height field $h(x,t)$ evolution over a sinusoidally patterned substrate undergoing athermal 
    (panel (a)) and fluctuating (panel (b)) dewetting, respectively. In panel (c) we report the contact angle profile $\theta(x)$ (Eq.~(\ref{eq:sine_angle})).} 
    \label{fig:patterned_sine8_difference_20-30}
\end{figure}

The sine wave pattern is defined by
\begin{equation}\label{eq:sine_angle}
    {\theta^{(1)}}(x) = \frac{5\pi}{36} + \frac{\pi}{36} \sin\left(q_{\theta} x\right),
\end{equation}
where the wavenumber is set to $q_{\theta} = 2\pi/512$ (thus the wavelenght is $\lambda_{\theta}=512$), such that 
the contact angle ranges between $\pi/9$ ($20^{\circ}$) and $\pi/6$ ($30^{\circ}$).
\textcolor{black}{We plot the space-time evolution of the height field on this kind of patterned substrate in Figs.~\ref{fig:patterned_sine8_difference_20-30}(a) (deterministic) and  \ref{fig:patterned_sine8_difference_20-30}(b) (stochastic). 
Both show, as expected, that the film starts to rupture prevelently in regions of high contact angle and droplet nucleation occurs in regions of low contact angle. Indeed, the deterministic dewetting leads to the formation of precisely $L/\lambda_{\theta}$ droplets.}
\textcolor{black}{However, in the stochastic case, the morphology of the dewetting process seems to be 
less bound to the subjacent pattern: we report, indeed,
the formation of few droplets ($\approx 0.2 \, (L/\lambda_{\theta})$, on average) in regions of high contact angle ($\theta(x) \approx \pi/6$).
}
We also notice that thermal fluctuations are able to induce defects in the \textit{stable} droplet state, i.e. in regions of low contact angle ($\theta(x) \approx \pi/9$). 
This is shown by the double droplet state in panel \ref{fig:patterned_sine8_difference_20-30}(b) at $x \approx 3 \lambda_{\theta}$ and happens in about 20\% of our stochastic simulations.
\textcolor{black}{Concerning the time scales involved, it has to be stressed that, owing to the contact angle inhomogeneity, a characteristic time as  
in Eq.~(\ref{eq:t0}) is not anymore uniquely defined. Still, it is natural to assume that the most unstable regions of the substrate, where the contact angle is the highest ($\max_x(\theta(x)) = \pi/6$), are the main drive to the dewetting and, hence, determine the relevant time scales. We set, therefore, $t_0 \equiv t_0(\pi/6)$. With this choice, we observe that the film starts to rupture at around 
$t \equiv \tau_r \gtrsim 2 t_0$ in the derministic simulation, which is indeed comparable with the homogeneous substrate 
(see Fig.~\ref{fig:rupture_times_semilogy_more_theta}). By analogy we take as the reference wavenumber $q_0 = q_0(\pi/6)$. Consequently, since 
$q_0$ decreases with $\theta$, we should expect the actual most unstable wavenumber to be slightly below $q_0$.}\\
\begin{figure}
    \centering
    \includegraphics[width=0.48\textwidth]{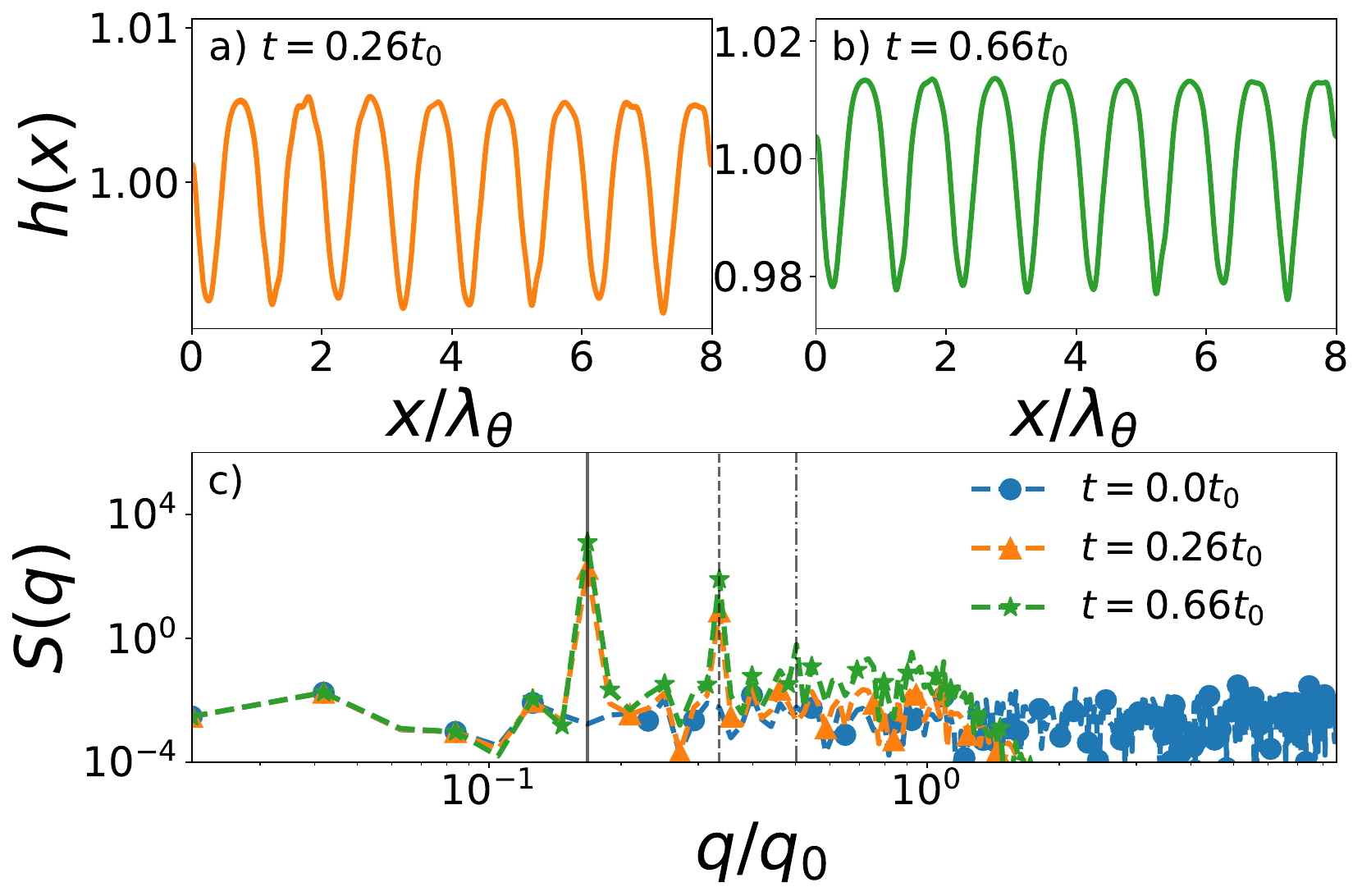}
    \caption{\textcolor{black}{(color online)} Height profiles from deterministic dewetting ($\sigma=0$) 
    over the sinusoidally patterned substrate, Eq.(\ref{eq:sine_angle}) at $t=0.26 t_0$ (panel a)) and $t=0.66 t_0$ (panel b))  
    and corresponding spectra (panel c)).  
    In panel c) also the initial spectrum, $S_0(q)$, is reported; the vertical lines indicate the pattern wavemode $q_{\theta}$ (solid) and its multiples $2q_{\theta}$ (dashed) and $3q_{\theta}$ (dashed dotted).
    As explained in the text, the convention $t_0 = t_0(\pi/6)$ and $q_0 = q_0(\pi/6)$ applies.}   
    \label{fig:spectral_analysis_deter_sine8}
\end{figure}
\begin{figure}
    \centering
    \includegraphics[width=0.48\textwidth]{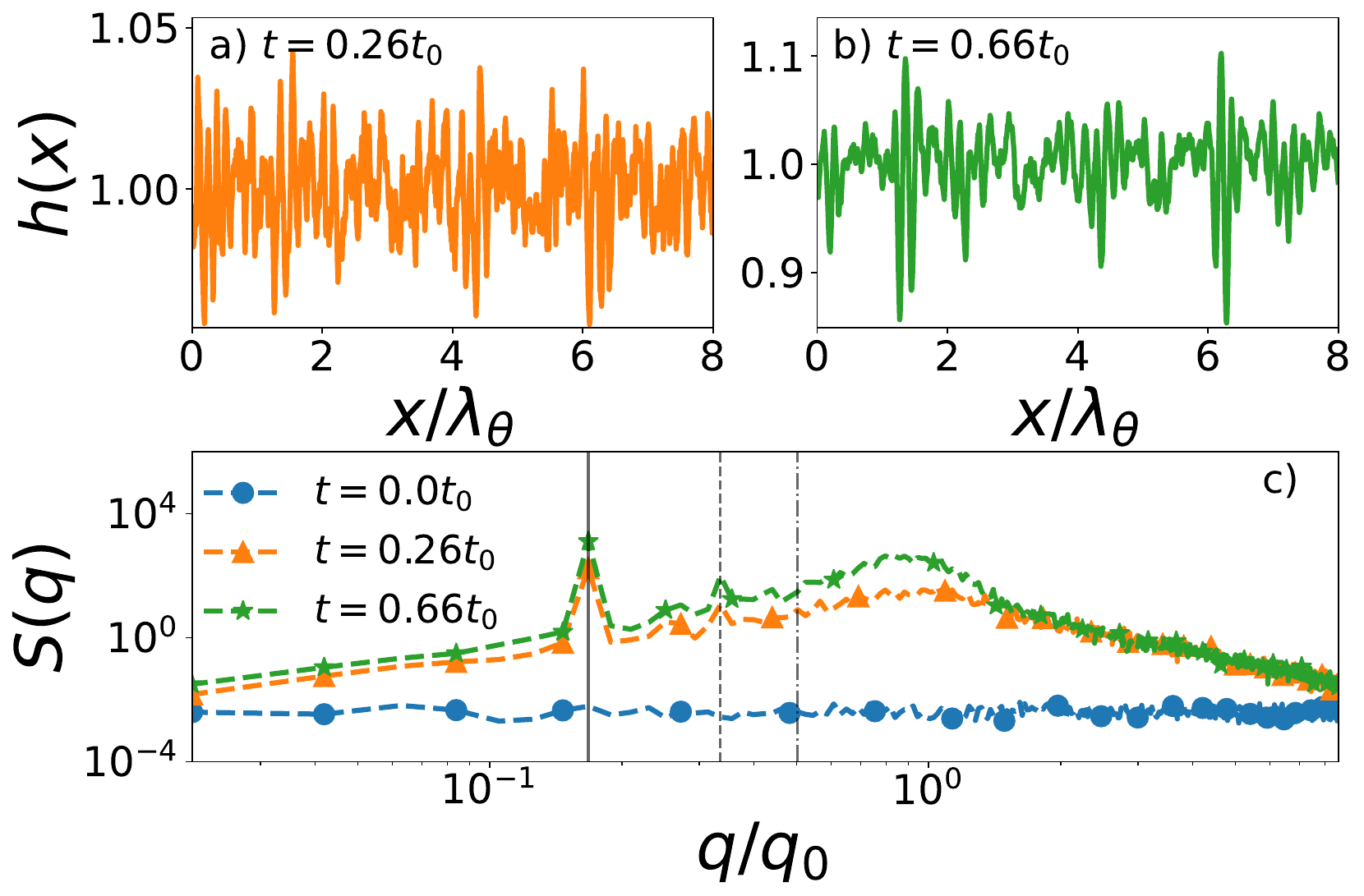}
    \caption{\textcolor{black}{(color online)} Height profiles from stochastic dewetting ($\sigma=\sigma_0$) 
    over the sinusoidally patterned substrate, Eq.(\ref{eq:sine_angle}) at $t=0.26 t_0$ (panel a)) and $t=0.66 t_0$ (panel b))  
    and corresponding spectra (panel c)).  
    In panel c) also the initial spectrum, $S_0(q)$, is reported; the vertical lines indicate the pattern wavemode $q_{\theta}$ (solid) and its multiples $2q_{\theta}$ (dashed) and $3q_{\theta}$ (dashed dotted).}
    \label{fig:spectral_analysis_stoch_sine8}
\end{figure}
The influence of the substrate wettability modulation shows up already in the early stages of dewetting ($t<t_0$), as reflected
 in the shape and evolution of the spectra (Figs.~\ref{fig:spectral_analysis_deter_sine8}-\ref{fig:spectral_analysis_stoch_sine8}). 
The maximum of the spectrum from the athermal system 
(Fig.~\ref{fig:spectral_analysis_deter_sine8}(c)), in fact, is not located at $q \approx q_0$ anymore.
Instead, the wavenumber of the substrate pattern sets the (absolute) maximum at $q=q_{\theta}$. 
Interestingly, further local maxima (of progressively decreasing amplitude) can be detected at integer multiples of $q_{\theta}$, namely for $q=2q_{\theta}$ and $q=3q_{\theta}$ (indicated, in the figure, by the gray dashed and dashed-dotted lines, respectively).
For higher values of $q$, the spectrum tends to that of spinodal dewetting,
with a small local maximum at $q \lesssim q_0$ (consistently with the above discussion about the choice of $q_0$)
and a fast decay for $q/q_0 > \sqrt{2}$.
The strong correlation between the patterning ($\theta(x)$) and the height field evolution may be further highlighted by visual inspection of the profiles in Fig.~\ref{fig:spectral_analysis_deter_sine8}(a-b), 
where a precise phase-shift of $\pi/2$, as compared to the contact angle profile,  Fig.~\ref{fig:patterned_sine8_difference_20-30}(c), can be observed. Such
shift, mathematically, stems from the fact that the height is forced by the pressure gradient and, hence, by the gradient of the disjoining pressure, which contains the contact angle dependence as $\Pi \propto \cos(\theta(x))$.
In the fluctuating dewetting, instead, the stochastic dynamics shadows the correlation,
as it can be appreciated from Figs.~\ref{fig:spectral_analysis_stoch_sine8}(a-b), where 
the height field is reported. 
Correspondingly, we see in the spectra, Fig.~\ref{fig:spectral_analysis_stoch_sine8}(c), 
that the "spinodal" maximum at $q \lesssim q_0$ is much more enhanced, than it was in the athermal case, and it is of comparable to the "pattern" maximum at $q = q_{\theta}$.
Furthermore, the local maxima at $n q_{\theta}$ are basically lost in the spinodal background.
\subsubsection{Square wave pattern}\label{subsubsec:square_wave}
\begin{figure}
    \centering
    \includegraphics[width=0.48\textwidth]{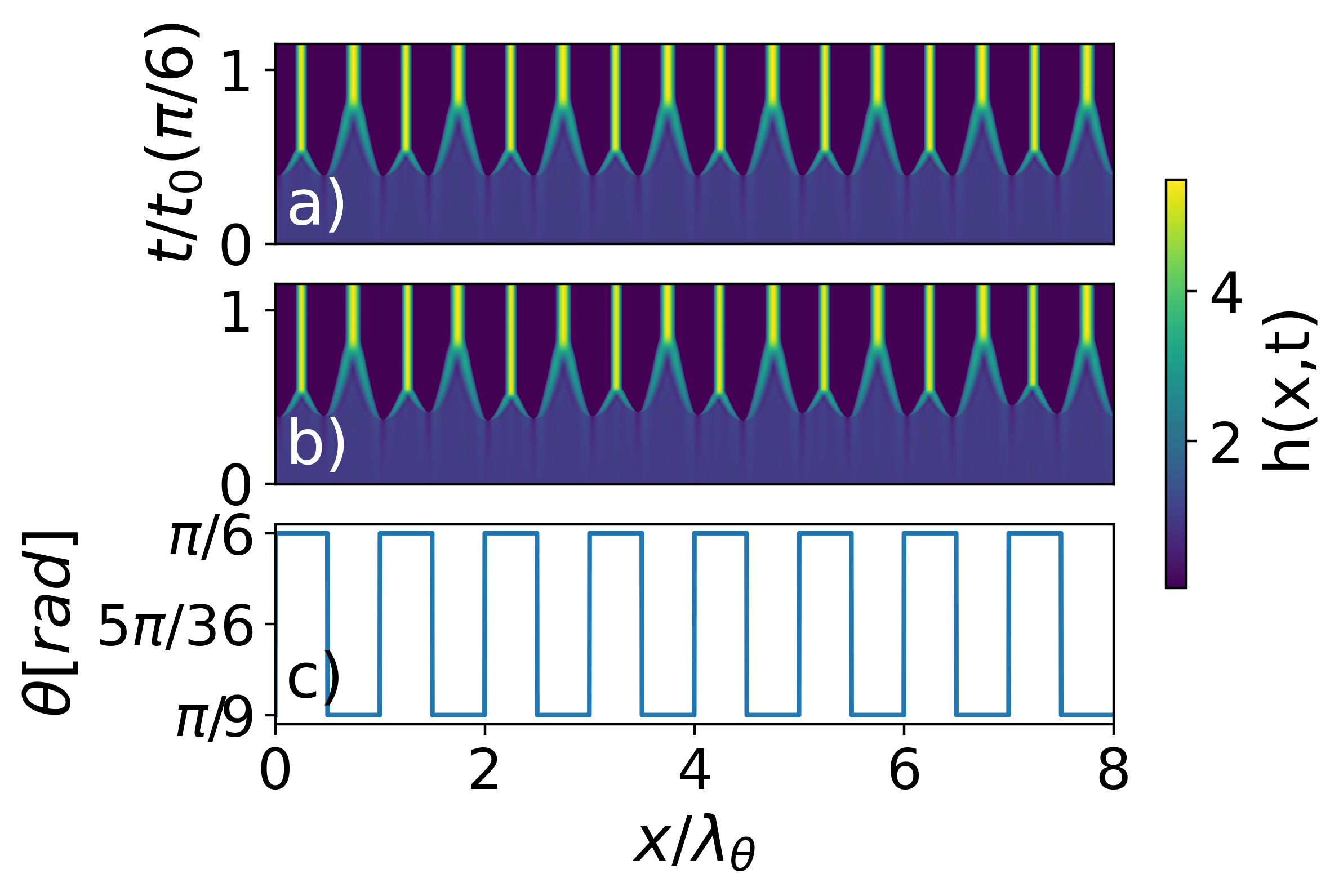}
        \caption{\textcolor{black}{(color online)} Space-time plot of the height field $h(x,t)$ evolution over a square-wave patterned substrate undergoing athermal 
    (panel (a)) and fluctuating (panel (b)) dewetting, respectively. In panel (c) we report the contact angle profile $\theta(x)$ (Eq.~(\ref{eq:sharp_contact_angle_spatial})).} 
    \label{fig:patterned_step8_difference_20-30}
\end{figure}
In the following we focus on the dewetting process on a substrate with a contact angle pattern given by 
\begin{equation}\label{eq:sharp_contact_angle_spatial}
    {\theta^{(2)}(x)} = 
    \left\{
    \begin{array}{ll}
        \frac{\pi}{9} & \text{if} \quad\sin(q_{\theta}x) \leq 0, \\
        \frac{\pi}{6} & \text{otherwise}
    \end{array}
    \right..
\end{equation}
Looking at Fig.~\ref{fig:patterned_step8_difference_20-30}, one can immediately notice two main differences in comparison to the sinusoidal pattern (cf.~Fig.~\ref{fig:patterned_sine8_difference_20-30}). The first one is that the impact of thermal fluctuations on the 
global dewetting morphology looks much weaker; in particular, in both deterministic and stochastic 
simulations, the film ruptures exactly at the wettability discontinuities.
Secondly, stable droplets are formed also in regions of high contact angle. 
As a consequence, a total amount of $2\frac{L}{\lambda_{\theta}}$ droplets are observed, which is twice as many as for the pattern $\theta^{(1)}(x)$.\\
The characteristic time scale of formation process, however, is not the same for all droplets, 
since it depends on the local contact angle.
We observe, indeed, that, droplets nucleate faster in regions of higher contact angle. 
Specifically, if we define the droplet formation time as the delay between the, $\theta$-dependent, droplet 
nucleation time, $t_d(\theta)$, and the film rupture time, $\tau_r(\theta)$, i.e.
\begin{equation}\label{eq:taud}
\tau_d(\theta) = t_d(\theta) - \tau_r(\theta), 
\end{equation}
and measure the ratio
\begin{equation}\label{eq:time_ratio_delta8_substrate}
    \Xi = \frac{\tau_d(\pi/6)}{\tau_d(\pi/9)} 
\end{equation}
of formation times over patches of different contact angle, we find $\Xi \approx 0.3$.
Theoretically, this can be explained according to a retracting film scenario. We expect, in fact, that 
the characteristic time (\ref{eq:taud}) will be inversely proportional to the speed of the 
receding contact line over a particular substrate patch, namely $t_d \propto U^{-1}$; such speed, in turn,
is known to depend on the contact angle as $U(\theta) \sim \theta^3$ \cite{PhysRevE.82.056314}, 
therefore we get for the ratio $\Xi$:
\begin{equation}\label{eq:ratio_U_theta_qubed_1/3}
  \Xi = \frac{\tau_d(\pi/6)}{\tau_d(\pi/9)}  \propto  {\frac{U(\pi/9)}{U(\pi/6)} \sim \frac{\left(\pi/9\right)^3}{\left(\pi/6\right)^3} = \left(\frac{2}{3}\right)^3 = 0.296...},
\end{equation}
in excellent agreement with the measured value.\\
\begin{figure}
    \centering
    \includegraphics[width=0.48\textwidth]{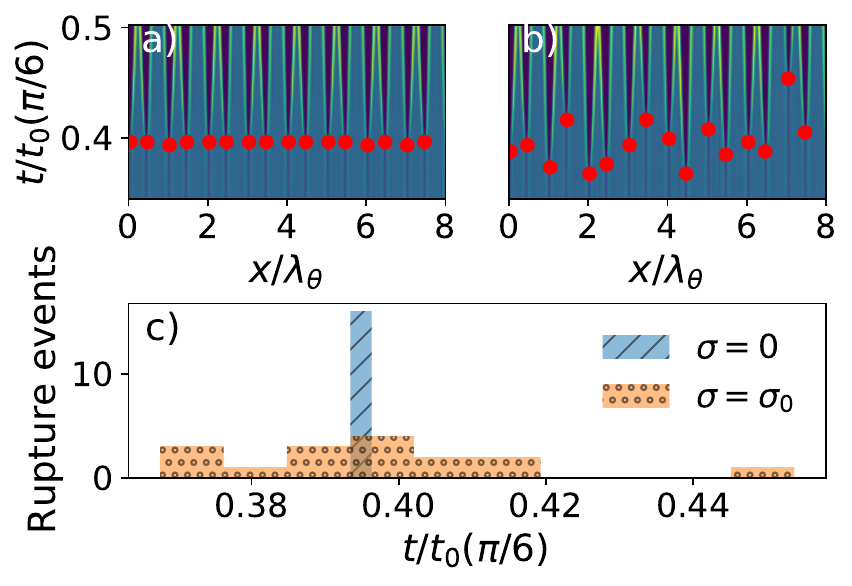}
    \caption{\textcolor{black}{(color online)} Panel a) and b): Space-time plots showing the evolution of the height field for athermal, a), and fluctuating, b), dewetting over the square-wave patterned substrated, 
    in a neighbourhood of the instant of time at which the first rupture event (in the athermal case) occurred; for the sake of visualization we mark the rupture events with red bullets (\textcolor{red}{$\bullet$}).
    Panel c): Distribution of times of occurrence of rupture events for the athermal ($\sigma=0$, blue, line-patterned) and fluctuating ($\sigma = \sigma_0$, orange, dot-patterned) dewetting.} 
    \label{fig:rupture_time_distri_square_wave8}
\end{figure}
The rupture times from the athermal and fluctuating dewetting films are shorter than 
their counterparts on the $\theta^{(1)}(x)$ pattern (Eq.~(\ref{eq:sine_angle})), and, unlike those, they are comparable to each other, hinting at a stronger bond to the substrate modulation (also in the fluctuating case).
However, a difference arises in the full time distributions of rupture events, 
shown, as histograms, in
Fig.~\ref{fig:rupture_time_distri_square_wave8}(c) 
(in Fig.~\ref{fig:rupture_time_distri_square_wave8}(a) and 
Fig.~\ref{fig:rupture_time_distri_square_wave8}(b) we highlight the rupture events with red bullets in 
the time-space evolution diagram).
The blue (orange) bars are the data from the deterministic (stochastic) simulation.
We see that, while in the athermal case rupture events are concentrated in a very narrow time frame 
(i.e. they all occur almost simultaneously along the whole domain),
the distribution is broadened when thermal fluctuations are switched on 
(analogously to what happened with the droplet height distributions over the unstructured substrates).
Here, the rupture events are scattered over a time frame of $0.08\,t_0$ as compared to $0.003\,t_0$ 
in the deterministic simulation.\\
 \begin{figure}
     \centering
     \includegraphics[width=0.48\textwidth]{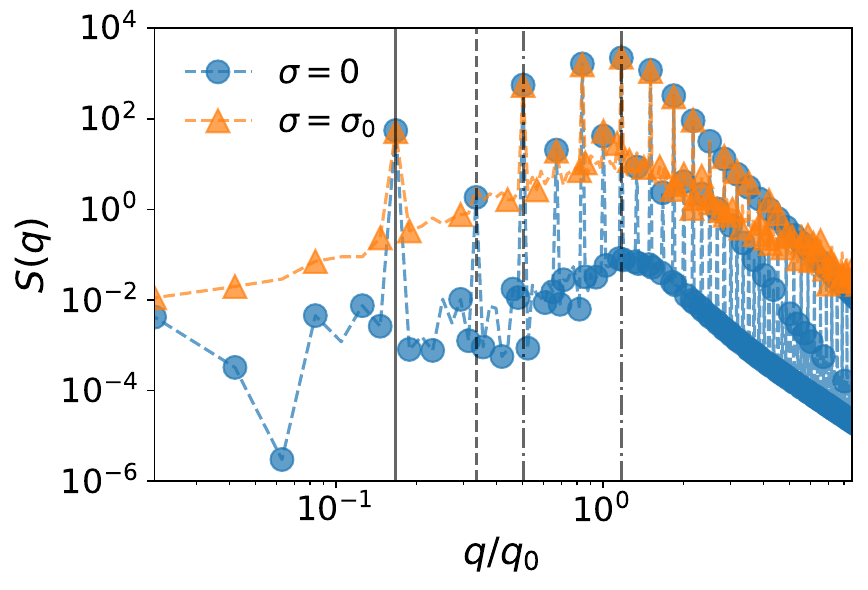}
     \caption{\textcolor{black}{(color online)} Spectra of the dewetting film on the $\theta^{(2)}(x)$ pattern, Eq.(\ref{eq:sharp_contact_angle_spatial}), at $t = 0.1t_0(\pi/6)$, with ($\sigma = \sigma_0$, \textcolor{pyorange}{$\blacktriangle$}) and without ($\sigma =0$, \textcolor{pyblue}{$\bullet$}) thermal fluctuations. 
     The vertical lines indicate the substrate wavemode $q_{\theta}$ (solid) and its multiples $2q_{\theta}$ (dashed), $3q_{\theta}$ (thin dashed-dotted) and $7q_{\theta}$ (thick dashed-dotted).}
     \label{fig:square_wave_both}
 \end{figure}
 As a final analysis we discuss the short time (pre-rupture) spectra on this substrate as well (see Fig.~\ref{fig:square_wave_both}).
 We show the deterministic (blue dashed line with bullets) and stochastic (orange dashed line with triangles) data at the same time $t=0.1\,t_0$.
 Under the light of what obtained with the sine wave pattern and given the dewetting morphology shown in Fig.~\ref{fig:patterned_step8_difference_20-30}, we would guess to find the absolute maximum of the spectrum at $q=2 q_{\theta}$.
Surprisingly, instead, the figure shows that, although a local maximum located at $2q_{\theta}$ can 
indeed be detected, there are several higher peaks at $n q_{\theta}$, with a global maximum at $7q_{\theta} \approx 1.2 q_0$; this probably suggests that the early time dewetting on this kind of substrate is still reminiscent of the spinodal background. Nevertheless, the strong bond of the dynamics with the
 pattern is evident and it is further pointed out by the unexpected observation that corresponding maxima
 from the deterministic and stochastic spectra are all of the same magnitude.

\section{Summary and Conclusions}\label{sec:sum_conclu}

\textcolor{black}{We have presented a lattice Boltzmann method for the simulation of the stochastic thin-film equation. 
The method has been tested against exact results on the height fluctuations spectrum of a dewetting film.  
We observed, in agreement with previous studies, that the inclusion of thermal fluctuations accelerates the dewetting and reduces 
the rupture times \cite{Grun2006}.
Furthermore, we have shown that the distribution of droplet sizes measured in the stochastic simulations 
is more spread out than its deterministic counterpart \cite{PhysRevE.92.061002}. \\
A central contribution of our work concerns the study of the role of the liquid-substrate interactions, parametrised by the contact angle $\theta$, on the dewetting process, with and without thermal fluctuations.}
\textcolor{black}{We reported and justified theoretically that the ratio of the deterministic and stochastic rupture times
decreases monotonically with the contact angle, though being always larger than one
(i.e. the fluctuating dewetting occurs faster than the athermal one, irrespective of the substrate wettability).  
We then performed simulations with a chemically patterned substrate, modelled as a space-dependent contact angle $\theta(x)$. 
Depending on the pattern, a smoothly varying profile (a sine wave) and a sequence of
alternate stripes (segments, in $1$-d), one can control the number of droplets formed,
which differ by a factor two, in spite of the two patterns having the same ``wavelength''.
In both cases, on the other hand, the dynamics appears strongly enslaved to the wettability 
modulation, so much that the effect of thermal fluctuations is significantly hindered. 
This even holds in the early stages of dewetting, as inferable from the inspection of the spectra.
Before concluding, let us make two last remarks. Firstly, it is worth mentioning that we do not 
expect that the results presented would differ too much in two dimensions (i.e. for real substrates).
The physical observables discussed, in fact, pertain essentially to the early dewetting (spectra, rupture times, droplet size distributions),
whereby the dynamics is mostly determined by the spectral properties of the linearized equations and by the characteristics of the free energy.
The signature of the dimensionality should, instead, emerge in the morphology of the dewetting pattern, especially in the long time evolution,
when coarsening and droplet coalescence dominate. These aspects will be subject of a forthcoming study. 
Secondly, we underline that the versatility of the numerical scheme would allow, in principle, to extend the method to simulate
the dynamics of thin films of more complex liquids, such as non-Newtonian and active fluids~\cite{Eggers1997,Carenza2019}.
}

\begin{acknowledgements}
The authors acknowledge financial support by the Deutsche Forschungsgemeinschaft (DFG) within the Cluster of Excellence ``Engineering of Advanced Materials'' (project EXC 315) (Bridge Funding) and the priority program SPP2171 ``Dynamic Wetting of Flexible, Adaptive, and Switchable Substrates'', within project HA-4382/11. 
The work has been partly performed under the Project HPC-EUROPA3 (INFRAIA-2016-1-730897), with the support of the EC Research Innovation Action under the H2020 Programme; in particular, S. Z. gratefully acknowledges the support of Consiglio Nazionale delle Ricerche (CNR) and the computer resources and technical support provided by CINECA.
We thank Paolo Malgaretti, Massimo Bernaschi and Mauro Sbragaglia for fruitful discussions.
\end{acknowledgements}


%

\end{document}